\documentclass[12pt]{article}

\begin{document}

\baselineskip 0.75cm
\topmargin -0.6in
\oddsidemargin -0.1in

\let\ni=\noindent

\renewcommand{\thefootnote}{\fnsymbol{footnote}}

\pagestyle {plain}

\setcounter{page}{1}

\pagestyle{empty}

~~~

\begin{flushright}
IFT--01/21
\end{flushright}

\vspace{0.3cm}

{\large\centerline{\bf Sterile neutrino creating a reduced LSND 
effect{\footnote{Supported in part by the Polish State Committee 
for Scientific Research (KBN), grant 5 PO3B 119 20 (2001-2).}}}}

\vspace{0.3cm}

{\centerline {\sc Wojciech Kr\'{o}likowski}}

\vspace{0.3cm}

{\centerline {\it Institute of Theoretical Physics, Warsaw University}}

{\centerline {\it Ho\.{z}a 69,~~PL--00--681 Warszawa, ~Poland}}

\vspace{1.2cm}

{\centerline{\bf Abstract}}

\vspace{0.3cm}

Although the hypothetic sterile neutrino $\nu_s$ is probably not involved 
significantly in the deficits of solar $\nu_e$'s and atmospheric $\nu_\mu $'s,
 it may cause the possible LSND effect.
In fact, we face such a situation, when the popular nearly bimaximal texture 
of active neutrinos  $ \nu_e\,, \,\nu_\mu\,, \,\nu_\tau $ is perturbed through
 a small rotation in the 14 plane, where $\nu_4 $ is the extra neutrino mass 
state induced by the sterile neutrino $\nu_s $. Then, with $m^2_1\simeq m^2_2$ 
we predict in the simplest case of $s_{13} \rightarrow 0$ that 
$ \sin^2 2\theta_{\rm LSND} = s^4_{14}/2$ and 
$ \Delta m^2_{\rm LSND} = |\Delta m^2_{41}|$. However, the negative Chooz 
experiment imposes on $s^4_{14}/2$ the upper bound $1.3 \times 10^{-3}$, 
suggesting a reduction of the amplitude of possible LSND effect.

\vspace{0.6cm}

\ni PACS numbers: 12.15.Ff , 14.60.Pq , 12.15.Hh .

\vspace{0.8cm}

\ni June 2001

\vfill\eject

~~~
\pagestyle {plain}

\setcounter{page}{1}

\vspace{0.2cm}

As is well known, the existence of at least one sterile neutrino $ \nu_s $ is 
needed --- beside  three familiar active neutrinos 
$ \nu_e\,, \,\nu_\mu\,, \,\nu_\tau $ --- in order to explain in terms of 
neutrino oscillations not only the observed deficits of solar $\nu_e $'s 
and atmospheric $\nu_\mu $'s, but also the possible LSND effect for 
accelerator $\bar{\nu}_\mu$'s and $\nu_\mu $'s. In this note we consider 
the phenomenologically popular nearly bimaximal texture of three active 
neutrinos \cite{R1} describing nicely both deficits,

\begin{equation} 
U^{(0)} =  \left( \begin{array}{cccc} c_{13}/\sqrt{2} & c_{13}/\sqrt{2} 
& s_{13} & 0 \\
- (1+s_{13})/{2}\, & \;\,(1-s_{13})/2 & c_{13}/\sqrt{2} 
& 0 \\ \;\,(1-s_{13})/2 & -(1+s_{13})/2\, & c_{13}/\sqrt{2} 
& 0 \\ 0 & 0 & 0 & 1 \end{array} \right) \;,
\end{equation} 

\ni where $ c_{12} = 1/\sqrt{2} = s_{12}$ and 
$ c_{23} = 1/\sqrt{2} = s_{23}$ with $ \delta_{13} = 0$, and then we 
perturb it through multiplying on the right by the rotation in the 14 
plane  with $ \delta_{14} = 0$. This results into the following 
four-neutrino texture:

\begin{equation} 
U =   \left( \begin{array}{cccc} c_{13}c_{14}/\sqrt{2} 
& c_{13}/\sqrt{2} & s_{13} 
& c_{13}s_{14}/\sqrt{2} \\ - (1+s_{13})c_{14}/{2}\, 
& \;\,(1-s_{13})/2 & c_{13}/\sqrt{2} 
& -(1+s_{13})s_{14}/2 \\ \;\,(1-s_{13})c_{14}/2 
& -(1+s_{13})/2\, & c_{13}/\sqrt{2} 
& \;\,(1-s_{13}) s_{14}/2 \\ -s_{14} & 0 & 0 & c_{14}
\end{array} \right) \;.
\end{equation} 

\ni Here, of course, $ c_{ij} = \cos \theta_{ij}$ and 
$ s_{ij} = \sin \theta_{ij}$. Such a four-neutrino mixing 
matrix can be obtained from its generic form by additionally 
putting $ s_{24} = 0 $ and $ s_{34} = 0 $. With  
$ U = \left( U_{\alpha i }\right)$ we can write

\begin{equation}
\nu_\alpha = \sum_i U_{\alpha i} \nu_i \;,
\end{equation}

\ni where $\nu_\alpha = \nu_e\,, \,\nu_\mu\,, \,\nu_\tau \,,\, \nu_s $ 
and  $\nu_i = \nu_1\,, \,\nu_2\,, \,\nu_3\,,\, \nu_4 $ are the flavor 
and mass neutrinos, respectively.

In the following we will discuss the consequences of the texture (2), 
showing that both $ s_{13}$ and $ s_{14}$ should be small to describe 
reasonably neutrino data ($ s_{13}$ might even vanish). In particular, 
the LSND effect, if confirmed, becomes a "\,$\!$sterile" perturbation 
of the nearly bimaximal texture (1) with the amplitude 
$ \sin^2 2\theta_{\rm LSND} = (1 + 2 s_{13}) s^4_{14}/2$   and mass-square 
scale $ \Delta m^2_{\rm LSND} = | m^2_4 - m^2_1|$, while the solar and 
atmospheric mass-square scales are $ \Delta m^2_{\rm sol} = m^2_2 - m^2_1$ 
and $ \Delta m^2_{\rm atm} = m^2_3 - m^2_2 $, respectively, where 
$ m^2_1 \stackrel{<}{\sim} m^2_2 \ll m^2_3 $. Thus, the sterile neutrino 
$\nu_s $ is here needed only to create the possible LSND effect.

We start from the familiar formulae for probabilities of neutrino 
oscillations $\nu_\alpha \rightarrow \nu_\beta $ on the energy shell:


\begin{equation}
P(\nu_\alpha \rightarrow \nu_\beta) = 
|\langle \nu_\beta| e^{i PL} |\nu_\alpha  \rangle 
|^2 = \delta _{\beta \alpha} - 4\sum_{j>i} U^*_{\beta j} 
U_{\beta i} U_{\alpha j} U^*_{\alpha i} \sin^2 x_{ji} \;,
\end{equation}

\ni valid if the quartic product 
$ U^*_{\beta j} U_{\beta i} U_{\alpha j} U^*_{\alpha i} $ is real, 
what is certainly true when a possible CP violation can be ignored 
[then $U^*_{\alpha i} = U_{\alpha i} $, as in the case of Eq. (2), 
and $P(\nu_\alpha \rightarrow \nu_\beta)  = 
P(\nu_\beta \rightarrow \nu_\alpha) $]. Here,


\begin{equation} 
x_{ji} = 1.27 \frac{\Delta m^2_{ji} L}{E} \;\;,\;\;  
\Delta m^2_{ji} = m^2_j - m^2_i
\end{equation}

\ni with $\Delta m^2_{ji}$, $L$ and $E$ measured in eV$^2$, km and GeV, 
respectively ($L$ and $E$ denote the experimental baseline and neutrino 
energy, while $ p_i = \sqrt{E^2 - m_i^2} \simeq E -  m^2_i/2E $ are 
eigen\-states of the neutrino momentum $P$).

With the use of oscillation formulae (4), the proposal (2) for the 
four-neutrino mixing matrix leads, in particular, to the following 
probabilities if $ m^2_1 \simeq m^2_2$:


\begin{eqnarray} 
P(\nu_e\;\; \rightarrow \;\;\nu_e) & \simeq & 1 -  c^4_{13}c^2_{14} 
\sin^2 x_{21} - c^2_{13} \left(1+ c^2_{14}\right) 
\left( 2s^2_{13}\sin^2 x_{32} + c^2_{13} s^2_{14} 
\sin^2 x_{41}\right) \nonumber \\ & &  - 
\;2 c^2_{13} s^2_{13} s^2_{14} \sin^2 x_{43}\;, \nonumber  \\
P( \nu_{\mu,\tau} \rightarrow \nu_{\mu,\tau})\!\! 
& \simeq & 1 - \frac{1}{4}\, c^4_{13}c^2_{14} \sin^2 x _{21}\nonumber 
\\ & & - \left[\frac{1}{2}(1+s^2_{13})(1 + c^2_{14}) 
\mp s_{13}s^2_{14}\right] \!\left[ c^2_{13} 
\sin^2 x_{32}\! +\!\frac{1}{2}(1\!\pm\! 
s_{13})^2 s_{14}^2 \sin^2 x _{41}\right]\nonumber 
\\ & &  -  \frac{1}{2}c^2_{13}(1 \!\pm\! s_{13})^2 
s_{14}^2 \sin^2 x _{43} \;, \nonumber \\ 
P( \nu_{\mu,\tau} \rightarrow \;\;\nu_e)\! & \simeq & 
\frac{1}{2}\,c^4_{13}c^2_{14} \sin^2 x_{21} \nonumber \\ 
& & + c^2_{13}\left[ s^2_{14} \mp s_{13}(1 + c^2_{14})\right] 
\left[ \mp  s_{13} \sin^2 x _{32} + \frac{1}{2}
(1 \pm s_{13})s^2_{14}\sin^2 \! x _{41}\right] 
\nonumber \\ & & \pm c^2_{13} s_{13}(1\pm s_{13})s^2_{14} \sin^2 x_{43}\, . 
\end{eqnarray}

\ni Note also that $P(\nu_e\!\rightarrow \!\nu_s) = 
2 c^2_{13} c^2_{14}s^2_{14}\sin^2 x_{41}$. Here, 
$P(\nu_\alpha\! \rightarrow \!\nu_\beta) = 
P(\nu_\beta\! \rightarrow \!\nu_\alpha) $ as a possible PC violation 
is neglected.

For solar experiments \cite{R2}, under the assumption that 
$ (x_{21})_{\rm sol} \sim 1\ll (x_{32})_{\rm sol}$, 
$|(x_{41})_{\rm sol}|$, $|(x_{43})_{\rm sol}|$, the first Eq. (6) gives 


\begin{eqnarray} 
P(\nu_e \rightarrow \nu_e)_{\rm sol} & \simeq & 1 -  c^4_{13}c^2_{14} 
\sin^2 (x_{21})_{\rm sol}  - 2(c_{13}s_{13})^2 - 
\frac{1}{2}c^4_{13}(1+ c^2_{14})s^2_{14} \nonumber \\ 
& \simeq & 1 - c^2_{14} \sin^2 (x_{21})_{\rm sol} - 
\frac{1}{2}(1+ c^2_{14})s^2_{14} \,,
\end{eqnarray}

\ni where the second step is valid for $c^2_{13} \gg s^2_{13} \simeq 0 $. 
Treating Eq. (7) perturbatively with respect to the constant term 
proportional to $s^2_{14}$ (and neglecting $s^2_{13}$), we get in 
the zero perturbative order

\begin{equation}  
c^2_{14} =  \sin^2 2\theta_{\rm sol} \sim 0.66 
\;{\rm or}\;0.97 \;{\rm or}\;0.80 \;,\; \Delta m^2_{21} = 
\Delta m^2_{\rm sol} \sim  (10^{-5} \;{\rm or}\; 10^{-7} 
\;{\rm or}\; 10^{-10})\;{\rm eV}^2 
\end{equation}

\ni and so,

\begin{equation} 
\frac{1}{2}(1+ c^2_{14})s^2_{14} \sim 
0.28 \;\;{\rm or}\;\;0.030 \;\;{\rm or}\;\; 0.18 \;,
\end{equation}

\ni if the recent estimation \cite{R3} of two-flavor LMA or LOW or VAC solar 
solution is used, respectively. Thus, this treatment is reasonable 
for LOW solar solution. In contrast, for LMA and VAC solar solutions 
the constant term in Eq. (7), modifying the familiar two-flavor structure 
of oscillation formula, cannot be neglected and so, in our four-neutrino 
texture these solar solutions cannot be applied without some changes 
tending to introduce into solar solutions a third additive parameter.

In the case of atmospheric experiments \cite{R2}, under the assumption that 
$ (x_{21})_{\rm atm} \ll (x_{32})_{\rm atm} \sim 1 \ll |(x_{41})_{\rm atm}|$, 
$|(x_{43})_{\rm atm}|$, the second Eq. (6) implies 

\begin{eqnarray} 
P( \nu_{\mu} \rightarrow \nu_{\mu})_{\rm atm} & \simeq & 1 - c^2_{13}\left[ 
\frac{1}{2} (1+s^2_{13})(1 + c^2_{14})  - s_{13}s^2_{14}\right] 
\sin^2 (x _{32})_{\rm atm} \nonumber \\ & & \;\;\,- \frac{1}{2} 
\left[ 1 - \frac{1}{4}(1+ s_{13})^2 s_{14}^2\right](1 + s_{13})^2 
s^2_{14} \nonumber \\ & \simeq & 1 - \left[ \frac{1}{2} 
(1 + c^2_{14}) - s_{13}s^2_{14}\right] \sin^2 (x _{32})_{\rm atm}
\nonumber \\ & & \;\;\,- \left[ \frac{1}{8}(3 + c^2_{14}) + 
\frac{1}{2}s_{13}(1 + c^2_{14})\right] 
s^2_{14} \;, 
\end{eqnarray}

\ni the second step being valid for $ c^2_{13} \gg s^2_{13} \simeq 0$ 
in the linear approximation in $ s_{13}$. Then, in the zero perturbative 
order in the constant term proportional to $s^2_{14}$ (and neglecting 
$s^2_{13}$, but not $s_{13}$) we obtain

\begin{equation} 
\sin^2 2\theta_{\rm atm} = \frac{1}{2}(1+ c^2_{14}) - s_{13}s^2_{14} 
\sim 0.99 - 0.030 s_{13} \;,\; \Delta m^2_{32} = \Delta m^2_{\rm atm} 
\sim 3 \times 10^{-3}\;{\rm eV}^2
\end{equation}

\ni and so,

\begin{equation} 
\frac{1}{8}(3+ c^2_{14}) s^2_{14} + \frac{1}{2}s_{13}(1+ c^2_{14}) s^2_{14}  
\sim 0.015 + 0.030 s_{13} \;,
\end{equation}

\ni if the LOW solar solution is used for $ c^2_{14}$.

Eventually, for the LSND accelerator experiment \cite{R4}, 
under the assumption 
that $ (x_{21})_{\rm LSND} \ll (x_{32})_{\rm LSND}\ll 
|(x_{43})_{\rm LSND}| \simeq |(x_{41})_{\rm LSND}| \sim 1$ 
the third Eq. (6) leads to

\begin{eqnarray}
P( \nu_\mu \rightarrow \nu_e)_{\rm LSND} & \simeq & \frac{1}{2} 
c^2_{13}(1 + s_{13})^2 s^4_{14} \sin^2 (x_{41})_{\rm LSND} 
\nonumber \\ & \simeq & \frac{1}{2} (1 + 2s_{13}) s^4_{14} 
\sin^2 (x_{41})_{\rm LSND} \;,
\end{eqnarray}

\ni where the second step holds for $ c^2_{13} \gg s_{13}^2 \simeq 0$ in 
the linear approximation in $ s_{13}$. Hence, the prediction

\begin{equation} 
\sin^2 2 \theta_{\rm LSND} = \frac{1}{2}\,c^2_{13} (1+s_{13})^2 s_{14}^4 
\simeq \frac{1}{2} (1+2s_{13}) s_{14}^4 \sim 4.5 (1+2s_{13}) \times 10^{-4}
\end{equation} 

\ni follows for the amplitude of LSND effect, if the LOW solar solution  
\cite{R3} is accepted for $c^2_{14} $. Here, 
$\Delta m^2_{\rm LSND} = |\Delta m^2_{41}|$. 
Note, however, that the predicted value (14) of 
$\sin^2 2 \theta_{\rm LSND}$ lies below its existing LSND lower 
limit $8\times 10^{-4}$ at 99\% CL (when $s_{13} < 0.39$, what 
is consistent with $c^2_{13} \gg s^2_{13} \simeq 0$).

The assumptions on $ (x_{ji})_{\rm sol}$, $ (x_{ji})_{\rm atm}$ 
and $ (x_{ji})_{\rm LSND}$ used above to derive Eqs. (7), (10) and 
(13) are all valid if {\it either}

\begin{equation} 
m^2_1 \simeq m^2_2 \ll m^2_3 \ll m^2_4 \sim {\it e.g.}\;1\,{\rm eV}^2 
\end{equation} 

\ni {\it or}

\begin{equation} 
m^2_4 \ll m^2_1 \simeq m^2_2 \simeq m^2_3 \sim {\it e.g.}\;1\,{\rm eV}^2 
\end{equation} 

\ni with

\begin{equation} 
\Delta m^2_{21} \sim (10^{-5}\; {\rm or}\; 10^{-7}\; {\rm or}\; 
10^{-10})\,{\rm eV}^2\;\;,\;\;\Delta m^2_{32} \sim 3 
\times10^{-3}\,{\rm eV}^2 
\end{equation} 

\ni in both cases. Then, in both cases the LSND mass-square scale 
is $\Delta m^2_{\rm LSND} = |\Delta m^2_{41}| 
\sim \; {\it e.g.}\; 1\;{\rm eV}^2 $ (of course, this value may 
be somewhat modified together with the values of $ m^2_1$ and $ m^2_4$).

Finally, for the Chooz reactor experiment \cite{R5}, where it happens that 
$ (x_{ji})_{\rm Chooz}\simeq (x_{ji})_{\rm atm}$, the first Eq. (6) predicts

\begin{eqnarray} 
P(\bar{\nu}_e \rightarrow \bar{\nu}_e)_{\rm Chooz} & \!\simeq\! & 
P(\bar{\nu}_e \rightarrow \bar{\nu}_e)_{\rm atm} \nonumber \\ 
& \!\simeq\! & 1 - 2 c^2_{13} s^2_{13}(1 + c^2_{14}) \sin^2 
(x_{32})_{\rm atm} - \frac{1}{2} c^2_{13} ( 1 + s^2_{13} + 
c^2_{13} c^2_{14}) s^2_{14} \,,
\end{eqnarray} 

\ni since $\sin^2 (x_{41})_{\rm atm}= 1/2 $ and 
$\sin^2 (x_{21})_{\rm atm} \ll \sin^2 (x_{32})_{\rm atm}$ 
due to $| (x_{41})_{\rm atm}| \gg (x_{32})_{\rm atm} \sim 1 
\gg (x_{21})_{\rm atm}$ with $ |\Delta m^2_{41}| \gg 
\Delta m^2_{32}  \gg \Delta m^2_{21}$. For $ c^2_{13} 
\gg s^2_{13} \simeq 0$ the formula (18) is reduced to

\begin{equation} 
P(\bar{\nu}_e \rightarrow \bar{\nu}_e)_{\rm Chooz} 
\simeq 1 - \frac{1}{2} (1 + c^2_{14}) s^2_{14} \;,
\end{equation} 

\ni where  $ (1 + c^2_{14}) s^2_{14}/2 \sim 0.030$ if the LOW solar 
solution is used for $ c^2_{14}$. In terms of the effective two-flavor 
oscillation formula, Eqs. (18) and (19) imply 

\begin{eqnarray} 
\sin^2 2 \theta_{\rm Chooz} \equiv \frac{P(\bar{\nu}_e \rightarrow 
\bar{\nu}_e)_{\rm Chooz}-1}{\sin^2 x_{\rm Chooz}}\!\! & = & 
\!\! 4 c^2_{13} s^2_{13}(1 + c^2_{14}) \sin^2 (x_{32})_{\rm atm} + 
c^2_{13}(1 + s^2_{13} + c^2_{13} c^2_{14})s^2_{14} \nonumber \\ 
& \simeq & \!\! (1 + c^2_{14}) s^2_{14}\;,
\end{eqnarray} 

\ni the second step being valid for $ c^2_{13} \gg s^2_{13} \simeq 0$. 
Here, $ x_{\rm Chooz} = |(x_{41})_{\rm atm}| \gg 1$ and thus 
$\sin^2 x_{\rm Chooz} = 1/2$.

The negative result of Chooz experiment \cite{R5} excludes the disappearance 
process of reactor $\bar{\nu}_e$'s for $\sin^2 2 \theta_{\rm Chooz} 
\stackrel{>}{\sim} 0.1$, when $\Delta m^2_{\rm Chooz} 
\stackrel{>}{\sim} 0.1\,{\rm eV}^2$ is considered (then 
$\Delta m^2_{\rm Chooz} \gg \Delta m^2_{\rm atm} 
\sim 3 \times 10^{-3}\,{\rm eV}^2$ and so, $ x_{\rm Chooz} 
\gg x_{\rm atm} \sim 1$, leading consistently to 
$\sin^2 x_{\rm Chooz} = 1/2$). Thus, the nonobservation 
of Chooz effect in the above parameter range implies that 
$\sin^2 2 \theta_{\rm Chooz} \stackrel{<}{\sim} 0.1$, when 
$\sin^2 x_{\rm Chooz} = 1/2$. Then, Eq. (20) requires

\begin{equation} 
4 c^2_{13} s^2_{13}(1 + c^2_{14}) \sin^2 (x _{32})_{\rm atm} + 
c^2_{13}(1 + s^2_{13} + c^2_{13} c^2_{14}) s_{14}^2 \stackrel{<}{\sim} 0.1
\end{equation} 

\ni or for $ c^2_{13} \gg s^2_{13} \simeq 0$

\begin{equation} 
(1 + c^2_{14}) s_{14}^2 \stackrel{<}{\sim} 0.1\;.
\end{equation} 

\ni So, in the case of Eq. (22) there must be

\begin{equation} 
\frac{1}{2} s_{14}^4 \stackrel{<}{\sim}  1.3 \times 10^{-3}\;,
\end{equation} 

\ni what due to the definition of $\sin^2 2\theta_{\rm LSND}$ in Eq. (14) 
gives the upper bound

\begin{equation} 
\sin^2 2 \theta_{\rm LSND} \simeq \frac{1}{2}(1 + 2 s_{13}) s_{14}^4 
\stackrel{<}{\sim}  1.3 (1 + 2 s_{13}) \times 10^{-3}
\end{equation} 

\ni  for the amplitude of LSND effect. Note that this Chooz-induced 
bound for the LSND effect allows its amplitude to be equal to the value (14) 
predicted by the use of LOW solar solution. However, as was already mentioned,
 such magnitude (14) of $\sin^2 2 \theta_{\rm LSND}$ lies below its lower 
limit $8\times 10^{-4}$ following at 99\% CL from the existing LSND data 
(when $s_{13} < 0.39$, what is consistent with $ c^2_{13} \gg s^2_{13} 
\simeq 0$). Note also that the LMA and VAC solar solutions (in fact not 
applicable in our four-neutrino texture without changes), if used {\it on 
their face value}, lead to $\sin^2 2 \theta_{\rm LSND} 
\sim 5.8(1 + 2s_{13})\times 10^{-2}$ and $ 2.0(1 + 2s_{13})\times 10^{-2}$, 
respectively. These figures are excluded by the Chooz-induced bound (24), 
though they are not inconsistent with the existing LSND data.

If $ s_{13} \rightarrow 0$, our predictions (13) and (19) imply the amplitudes

\begin{equation} 
\sin^2 2 \theta_{\rm LSND} = \frac{1}{2}s^4_{14} \;\;,\;\; \sin^2 2 
\theta_{\rm Chooz} = (1 + c^2_{14}) s^2_{14}
\end{equation} 

\ni which together with the amplitude $\sin^2 2 \theta_{\rm sol} = c^2_{14}$ 
[as given in the zero perturbative order with respect to $s^2_{14}/c^2_{14}$ 
by Eq. (7)] lead to the sum rule

\begin{equation} 
\sin^2 2\theta_{\rm sol} + \frac{1}{2} \sin^2 2 \theta_{\rm Chooz} + \sin^2 
2 \theta_{\rm LSND} = 1
\end{equation} 

\ni for these three neutrino-oscillation amplitudes (each in the 
two-flavor approximation). This sum rule can $\!$be$\!$ derived$\!$ 
also$\!$ from$\!$ the$\!$ probability summation relation 
$\sum_\beta \!P(\nu_e\! \rightarrow\! \nu_\beta)$ $ =1$ 
(with $\beta = e\,,\, \mu\,,\, \tau\,,\, s$) considered under 
the assumption of $ m^2_1 \simeq m^2_2$ for solar $\nu_e $'s 
(when $|(x_{41})_{\rm sol}| \gg (x_{21})_{\rm sol} \simeq \pi /2 $). 
The sum rule (26) leaves room for the LSND effect, depending on the 
magnitude of the Chooz effect which is not observed yet.

In conclusion, when accepting the present Chooz results, we face in 
the framework of our four-neutrino texture the alternative: {\it either} 
there is no LSND effect at all (then $s_{14}= 0$, and we are left with 
the popular three-neutrino nearly bimaximal texture working very well for 
solar $\nu_e $'s and atmospheric $\nu_\mu $'s), {\it or} this effect exists
 but with an amplitude $\sin^2 2 \theta_{\rm LSND}\!\! = \!(1\! + \!2 s_{13}) 
s_{14}^4/2$ reduced due to the Chooz-induced upper bound 
$ 1.3 (1\! + \!2 s_{13})$ $ \times 10^{-3}$. If the two-flavor LOW solar 
solution is accepted for $c^2_{14}$, then $\sin^2 2 \theta_{\rm LSND} 
\sim 4.5 (1 + 2 s_{13}) \times 10^{-4}$. The two-flavor LMA and VAC solar 
solutions cannot be applied in our four-neutrino texture without  changes, 
because in our oscillation formula of solar $\nu_e $'s there appears a 
constant term, significant if these solutions are used. It spoils the 
familiar two-flavor structure of the oscillation formula. We hope that 
the Mini-BooNE experiment will be able to support or reject our simple 
four-neutrino texture, in particular its "\,$\!$sterile"\, perturbative 
aspect.

\vfill\eject

~~~~
\vspace{0.5cm}

\vfill\eject

\end{document}